\documentstyle[epsf,twocolumn,aps,pre]{revtex}
\renewcommand{\vector}[1]{{\bf #1}}
\newcommand{\etal}{\textit{et al.} }
\newcommand{\ie}{\textit{i.e.,}}
\newcommand{\eg}{\textit{eg.} }
\newcommand{\viz}{\textit{viz.}}

\newcommand{\half}{\frac{1}{2}}
\newcommand{\lessthanorequal}{\leq}
\newcommand{\greaterthanorequal}{\geq}
\newcommand{\definedas}{\equiv}
\newcommand{\muchlessthan}{\ll}
\newcommand{\aboutequal}{\simeq}
\newcommand{\expectation}[1]{\left\langle #1 \right\rangle}

\newcommand{\ket}[1]{\left| #1 \right\rangle}
\newcommand{\greaterthanorabout}
           {\mathrel{\raise.3ex\hbox{$>$\kern-.75em\lower1ex\hbox{$\sim$}}}}
\newcommand{\vol}[1]{{\bf #1}}

\newcommand{\?}[1]{}
\newcommand{\artitle}[1]{}
\long\def\omitt#1{}

\begin{document}
\author{Alexei V. Tkachenko and Thomas A. Witten\\
 \it The James Franck Institute, \\
\it The University of Chicago, Chicago, Illinois 60637 }
\date{submitted to {\sl Physical Review E}; typeset \today}
\title{\bf  Stress propagation through frictionless granular material}
\maketitle
\begin{abstract}
We examine the network of forces to be expected in a static assembly of hard,
frictionless spherical beads of random sizes, such as a colloidal glass.  Such
an assembly is minimally  connected: the ratio of constraint equations to
contact forces approaches unity for a large assembly.   However, the bead
positions in a finite subregion of the assembly are underdetermined.  Thus to
maintain equilibrium, half of the exterior contact forces are determined by the
other half.  We argue that the transmission of force may be regarded as
unidirectional, in contrast to the transmission of force in an elastic
material.  Specializing to sequentially deposited beads, we show that forces on
a given buried bead can be uniquely specified in terms of forces involving more
recently added beads.  We derive equations for the transmission of stress
averaged over scales much larger than a single bead.  This derivation requires
the {\it Ansatz} that statistical fluctuations of the forces are independent of
fluctuations of the contact geometry.  Under this {\it Ansatz}, the
$d(d+1)/2$-component stress field can be expressed in terms of a $d$-component
vector field.  The procedure may be generalized to non-sequential packings.  In
two dimensions, the stress propagates according to a wave equation, as
postulated in recent work elsewhere.  We demonstrate similar wave-like
propagation in higher dimensions, assuming that the packing geometry has
uniaxial symmetry.
In macroscopic granular materials we argue that our approach may be
useful even though grains have friction and are not packed sequentially.
\end{abstract}
\section {Introduction}
	The nature of the forces within a static pile of grains has proven
more subtle than one might expect.  Such a pile is an assembly of many
hard, spheroidal bodies that maintain their positions via a balance of
gravitational forces and contact forces with their
neighbors\cite{Guyon.1990,Grassberger}.  On the one hand, determining these
forces is a prosaic equilibrium problem.  Since the number of grains is
large, the long-established notions of continuum solid mechanics appear
applicable.  On the other hand, a pile of grains or beads is not a
solid.   The forces between beads are more problematic than those between the
atoms of a conventional solid. These latter forces are smoothly varying on the
scale of the separation and they arise from a potential energy that includes
attraction.  The forces on a grain are different.  First, they vary sharply
with interparticle distance, and there is no attraction.
Second, the frictional part of a contact force is not determined by the
macroscopic positions of the grains.  Rather, it depends on the how each
contact was formed.  The resulting macroscopic behavior of the pile is also
clearly different from that of a conventional solid. Arbitrarily slight forces
can disturb the pile, so that the notion of stable equilibrium is suspect.
Despite these complexities, we expect the mechanics of a granular pile to be
universal.  Hard, round grains appear to form piles of the same nature
independent of their composition or detailed shape.  We are led to think of
these as nondeformable objects that exert normal forces of constraint and
transverse forces limited by Coulomb's static friction limit.
\par
 	Recently, a puzzling discovery has underlined the subtlety of the
forces in a  conical heap of poured sand\cite{load}.  The supporting force
under the center, where the pile is deepest, is not maximal.  Instead, the
maximal force occurs along a circle lying between the edge and the center
of the pile.  From this circle the force decreases to a {\it minimum} at
the center.  In order to explain this puzzling ``central dip", a number of
inventive approaches have been taken. Some\cite{Edwards.Mounfield} seek to
account for the minimum qualitatively by viewing the pile as a stack of
concentric ``wigwams", whose sloping sides support the load.
Others\cite{Goddard} have shown that the central dip is compatible with the
conventional continuum mechanics, in which the pile is viewed as a central
elastic zone flanked by an outer plastic zone which is at the  Coulomb
static friction limit.    A third
group\cite{Wittmer.Nature,Claudin.PRL.1997,Bouchaud.Cates.PRL} has argued
that granular material requires a new constitutive law, a homogeneous,
local, linear constraint on the stress arising from the packing geometry.
 We shall call it a ``null stress" law.  Their proposed law gives a
continuum mechanics as
simple as that of a liquid or a solid, yet different from either.  The
hallmark of this law
is the hyperbolic equations governing the transmission of forces.
Hyperbolic equations,
such as the wave equation, obey causality.  The wave at the present is
unaffected by the
future.  In the null-stress picture, the vertical direction plays the role
of time.
Accordingly, forces at a given point in the pile are only influenced by
forces above that
point.  Force {\it propagates} as in a traveling wave.  The transmission is
{\it
unidirectional}, in contrast with conventional continuum elasticity.  The
equations of
continuum mechanics are elliptical.  According to these, a contact within
the pile should
be influenced by all the forces above or below it, as sketched in Figure
\ref{elastic.vs}.
\par
 A separate approach has given indirect evidence for the unidirectional
transmission of forces.  Coppersmith's\cite{Coppersmith.etal} heuristic $q$
model aims to account for the point-to-point variability of the contact
forces.  It imposes a unidirectional prescription for determining the
downward forces from one grain in terms of the forces acting on it from
above.  Both this model and
refinements\cite{Socolar,Clement97} of
it yield an exponential falloff of probability for large forces.  This
exponential falloff  agrees well with measured
distributions\cite{Coppersmith.etal,Mueth,Rajdai96}.  This exponential falloff
contrasts with the Gaussian falloff expected for a heterogeneous elastic
solid.
\par
 	In this paper we grapple with the relationship between the
conventional elastic view and the newer null-stress picture.  Our work
builds on several recent studies of the relationship between the
connectivity of a structure and the force transmission in it.  Alexander's
recent review\cite{floppy.networks} has explored the nature of
unconstrained degrees of freedom in a minimally connected  or
isostatic\cite{Guyon.1990} network.  Ball and
Edwards\cite{Ball.Edwards.counting} have explored  force transmission
through minimally-connected networks, assuming a fixed co-ordination number
for all particles.  \omitt{replaced: these same unconstrained motions in
the context of regular lattices, accounting for spatial dimensionality,
aspherical beads, andfriction. } They have shown that such lattices can
have constitutive equations of the null-stress form.  Our main aim here is
to broaden the class of systems that must show unidirectional force
transmission, as required by the
null-stress picture.
\par
 We present the discussion of the problem of stress transmission in a
granular packing on several levels of generality. In sections
\ref{counting}--\ref{macroscopic} we focus on properties of the system of
frictionless, spherical beads.
 Possible experimental realizations are hard-sphere colloidal
dispersions\cite{VanderWerff,Watanabe.shear.thick} or weakly-deformed droplet
emulsions\cite{Weitz.Mason}.  First, in Section \ref{counting} we use general
counting arguments to show that such a packing is minimally coupled.  We then
relate this fact to the inadequacy of elastic description for such a system.
For a small subsystem within the pile a counting of equations and unknowns
shows that approximately half of the surface forces transmitted from outside
the subsystem are redundant.  In equilibrium these cannot be independently
specified, but have a fixed relation to the other half of the forces.  This
requirement for balance amongst many forces implies the existence of soft
modes---infinitesimal deformations with no restoring force.  In the continuum
limit, the soft modes impose conditions on the stress field of the null-stress
form.
\par
In order to obtain the particular form of such macroscopic description, we
limit further discussion to so-called sequential packing specified in
Section
\ref{sequential}. In Section \ref{macroscopic} we give a microscopic
prescription for determining the contact forces in a unidirectional way, and
develop a green's-function formalism for determining the forces in the
$d$--dimensional  sequential packing.  This picture allows one to decouple the
geometric features of the packing from the pattern of transmitted forces.  In
Section \ref{macroscopic} we also explore the macroscopic consequences of these
force laws, leading to an expression for the stress tensor with $d$ variables
rather than the
$d(d+1)/2$ variables of a general stress tensor.   In two dimensions, our
formalism  places a constraint on the stress, whose form agrees with the null
stress law of Wittmer \etal
\cite{Wittmer.Nature}.  In higher dimensions, the constraints on the stress
also lead to a unidirectional equation for the transmission of stress in the
form of a wave equation. In Section 6 we consider the relevance of our findings
for real granular piles, which are not sequentially packed and which have
friction.  Friction can alter the transmission of forces qualitatively,
restoring the elastic transmission of stress.

\section{Underdetermination within a frictionless  pack}
\label{counting}
 In this section we consider how the constraints inherent in the packing of
impenetrable,  frictionless beads determine the contact forces between the
beads.   Forces in frictionless packs have been studied by
simulation\cite{Lacasse.Grest.Levine,Langer.Liu,Luding.frictionless,Farr}.
Theoretical properties of the forces have been established for simplified
systems
\cite{Ball.Edwards.counting,Moukarzel.PRL,Oron.Herrmann}. We begin with a
summary of the well-recognized enumeration of equations and unknowns, as
discussed, \eg by Alexander\cite{floppy.networks} and by the Cambridge
group\cite{Ball.Edwards.counting}.  We then consider the role of external
forces acting on a subregion of the pack, and show that a subset of these
suffices to determine the others.

For definiteness we consider a system of rigid spherical beads  whose sizes are
chosen randomly from some continuous distribution. We suppose that  the
diameters  of all the spheres are specified and the topology of their packing
is  given.  That is, all  the pairs of   contacting beads are specified. The
topology can be characterized by the average coordination number, \ie the
average number of nearest neighbors, $\overline Z =2N_c/M$ (here $N_c$ is the
total number of contacts, and $M$ is the number of beads). The necessary
condition for the packing of a given topology to be realizable in a space of
dimensionality $d$ is that the coordinates of the bead centers, ${\bf
x}_\alpha$  satisfy the following  equations,  one for each pair of contacting
beads $\alpha$ and $\beta$:
\begin{equation} \label{beadaverage}
\left({\bf x}_\alpha-{\bf x}_\beta\right)^2=\left(R_\alpha+R_\beta\right)^2
\end{equation}
Here $R_\alpha$, $R_\beta$ are the radii of the beads. There are $N_c$ such
equations (one for each contact), and $Md$ variables ($d$ for each bead).
The number of equations should not exceed the number of variables;
otherwise, the co-ordinates $\vector x_\alpha$ are overdetermined. Thus,
the coordination number of a geometrically-realizable packing should not
exceed the critical value of $2d$: $\overline{Z}\lessthanorequal 2d$.  We
assume all the equations imposed by the topological constraints to be
independent. If they were not independent,  they would become so upon
infinitesimal variation of the bead sizes. For instance, the hexagonal
packing in two dimensions has the coordination number 6 which is higher
then the critical value, $\expectation{ Z}=4$; but the extra contacts are
eliminated by an infinitesimal variation of the bead diameters. In other
words, the creation of a contact network with coordination number higher
than $2d$  occurs with probability zero in an ensemble of spheres with a
continuous distribution of diameters.  We shall ignore such zero-measure
situations henceforth.
\par
  The above consideration gives the upper limit on the average coordination
number, $\overline{Z}$. The lower limit can be obtained from the analysis
of mechanical stability of the packing: it gives a complementary
inequality: $\overline{Z} \greaterthanorequal 2d$.
We will consider a packing to be mechanically stable if there is a non-zero
measure set of external forces which can be balanced by inter-bead ones.
The packing of frictionless spheres is always characterized by
$\expectation{ Z} =2d$  , as we now show.
 Stability requires that the net force on each bead be zero; there are $Md$
such equations.  The forces in these $Md$ equations are the $N_c$ contact
forces.  The $Md$ equilibrium conditions determine the magnitudes of the
$N_c$ contact forces.  (Their directions are determined by the geometry of
the packing.)  The number of equilibrium equations $Md$ should not exceed
the number of force variables $N_c$; otherwise these forces would be
overdetermined.  Thus $Md \lessthanorequal N_c$, or
$\overline Z \greaterthanorequal 2d$.  To avoid both overdetermined
co-ordinates
and overdetermined forces, we must thus have $\overline Z = 2d$.
\par
 Similar counting arguments have been  discussed
previously\cite{floppy.networks,Moukarzel.PRL}.  A subset of them have been
applied to granular packs with friction\cite{Ball.Edwards.counting}. Here we
emphasize a further feature of a frictionless bead pack that has not been well
appreciated: the co-ordinates and forces within a subregion of a large bead
pack are necessarily
{\it underdetermined}.  Quantifying this  indeterminacy will play an important
role in our reasoning below.  To exhibit the indeterminacy, we consider some
compact region within  the packing, containing $M'$ beads.  This unbiased
selection of beads must have the same average co-ordination number $\overline
Z$ as the system as a whole: $\overline Z' = 2d$.  Let $N_{\rm ext}$ be the
number of contacts of this sub-system with external beads, and $N_{\rm int}$ be
the number of the internal contacts. The average coordination number
$\overline{Z}'$ can be expressed $\overline{Z}'=(N_{\rm ext}+2N_{\rm int})/M'$
(any internal contact should be counted twice). Since there are $M' d$
equations of force balance for these beads, one is able  to determine all
$N_{\rm ext}+N_{\rm int}$ contact forces in the system, whenever  $M'd = N_{\rm
ext} + N_{\rm int}$.
 Evidently, if the forces on the  $N_{\rm ext} $ contacts are not
specified, the internal forces cannot be computed: the system is
underdetermined.  The number of external forces $N_0$ required is given by
$N_0=M'd - N_{\rm int}$.  This $N_0$ may be related to the average
co-ordination number $\overline{Z}'$:
\begin{equation}
 N_0 = M'\left[ d - {\overline{Z}' \over 2} \right] + \frac{N_{\rm ext}}{2}
\end{equation}
\par
 We now observe that the quantity in $[ ... ]$ vanishes on average.  This
is because the average of $\overline{Z}'$ for any subset of particles is
the same as the overall average.  There is no systematic change of
$\overline{Z}'$ with $M'$.  Thus if one half (on average) of
mutually-independent external forces is known (let us call them ``incoming"
ones), the analysis of force balance in the region enables one to determine
all the remaining forces, including the other half of external ones
(``outgoing").   We are free to choose the incoming contacts at will,
provided these give independent constraint equations.
\par
This observation supports the unidirectional, propagating stress picture,
discussed in the Introduction.  Indeed, one can apply the above arguments
to the slabs of the packing created by cutting it with horizontal surfaces.
In a given slab of material, we choose the forces from the slab above as
the incoming forces.  According to the preceding argument, these should
determine the outgoing forces transmitted to the slab beneath.   This must
be true provided that the constraints from the upper slab are independent.

 Such force transmission contrasts with that of a solid body, as emphasized
in the Introduction.  If a given set of forces is applied to the top of a
slab of elastic solid, the forces on the bottom are free to vary, provided
the total force and torque on the slab are zero.  Yet in our bead pack,
which appears equally solid, we have just concluded that stability
conditions determine all the bottom forces individually.
 In deducing this peculiar behavior, we did not exclude tensile forces; we
may replace all the contacts by stiff springs that can exert strong positive or
negative force, without altering our reasoning.   In this sense our result
is different from the recent similar result of Moukarzel\cite{Moukarzel.PRL}.
The
origin of the peculiar behavior lies in the minimal connectivity of the beads.
\par
In a subregion of the minimal network, the constraints can be satisfied with
no internal forces.  Moreover, numerous (roughly
$N_{\rm ext}/2$) small external displacements can be applied to the subregion
without generating proportional restoring forces.  We call these motions with
no restoring force ``soft modes".   If we replace these external
displacements with external
forces and require no motion, compensating forces must be applied elsewhere to
prevent motion of the soft modes.  If the applied forces perturb all the soft
modes, there must be one compensating force for each applied force to prevent
them from moving---on average $N_{\rm ext}/2$ of them.  The subregion is
``transparent" to external forces, allowing them to propagate individually
through the region.
\par
This transparent behavior would be lost if further springs were added to
the minimal network, increasing $\overline{Z}$.  Then the forces on a
subregion would be determined even without external contacts.  The addition
of external displacements would deform the springs, and produce
proportional restoring forces.  There would be no soft modes, and no
transparency to external forces.

 A simple square lattice of springs provides a concrete example of the
multiple soft modes predicted above.  Its elastic energy has the form
\begin{equation}
H = K \int {\rm d}x{\rm d}y  \left[ (u^{xx})^2 + (u^{yy})^2 \right]
\end{equation}
This functional does not depend on  $u^{xy}$, thus there are shear
deformations  ($u^{xx}=u^{yy}=0$)  which cost no elastic energy. This means
that the stress field should be orthogonal to any such mode, \ie
\begin{equation}\sigma^{ij}{u_o}^{ij}=0
\end{equation}
where ${u_o}^{xx}= {u_o}^{yy}=0$, and ${u_o}^{xy}$ is an arbitrary function
of $(x;y)$. The above equation implies that $\sigma^{xy}=0$, \ie the
principal axes of the stress tensor are fixed and directed along $x$ and
$y$. This provides a necessary closure for the  standard macroscopic
equation of force balance,
\begin{equation}
\partial^i \sigma^{ij}=f^j_{\rm ext}
\label{balance}
\end{equation}
here $\bf f_{\rm ext}$ is an external force. Since  $\sigma^{xy}=0$ the two
unknown components of the stress field, $\sigma^{xx}$ and $\sigma^{yy}$
propagate independently along the corresponding characteristics, $x=const$
and $y=const$:
\begin{equation}
\partial^x \sigma^{xx}=f^x_{\rm ext}\end{equation}
\begin{equation}
\partial^y \sigma^{yy}=f^y_{\rm ext}\end{equation}
\par
 The propagation of the solution along characteristics is a property of
hyperbolic problems  such as  wave equation. The above equations without
external force imply that each component of the stress tensor $\hat \sigma$
satisfies a wave equation of the form
\begin{equation}
\left ({\partial^2 \over \partial t^2} - {\partial^2 \over \partial s^2}
\right )\hat \sigma = 0
\end{equation}
where $t \definedas x+y$ and $s \definedas x-y$.
Thus, the fact that the original elastic energy has the soft modes results
in hyperbolic, rather than elliptic equations for the stress field. One has
now to specify the surface forces  (or displacements) at a single
non-characteristic surface---a line not parallel to $x$ or $y$--- in order
to determine the stress field in all the sample.
\par
A frictionless granular packing behaves like this example: they both are
minimally coupled; they both have soft modes; they both have unidirectional
propagation. In both examples only the surface of the sample stabilizes the
soft modes.
The above consideration of regular lattice can be easily extended to the
case of arbitrary angle between the characteristic directions, $x$ and $y$.
Instead of starting with a square lattice, we could have applied a uniform
$x-y$ shear, altering the angle between the horizontal and vertical
springs.  The reasoning above holds for this lattice just as for the
original square one.
\par
The nature of the soft modes in a disordered bead pack is less obvious than
in this lattice example.  We have not proven, for instance, that all the
forces acting on the top of a slab correspond to independent soft modes,
which determine the forces at the bottom.  Otherwise stated, we have not
shown that the soft modes seen in the microscopic displacements have
continuum counterparts in the displacement field of the region.  However,
the following construction, like the lattice example above, suggests that
the soft modes survive in the continuum limit
\par
To construct the pack, we place circular disks one at a time into a
two-dimensional vertical channel of width $L$.  (Such sequential packings
will figure prominently in the next section.)  Since the disks are of
different sizes, the packing will be disordered.  We place each successive
disk at the lowest available point until the packed disks have reached a
height of order $L$, as shown in Figure \ref{channel}.  We now construct a
second packing, starting from a channel of slightly greater width $L +
\delta$.  We reproduce the packing constructed in the first channel as far
as possible.  We use an identical sequence of disks and place each at the
lowest point, as before.   There must be a nonvanishing range of $\delta$
for which the contact topology is identical.  The motion of the wall over
this range is thus a soft mode.  As the side wall moves, the top surface
will move by some amount $\epsilon$, proportional to $\delta$.  Now,
instead of holding the side wall fixed, we exert a given force $f^x$ on it.
Likewise, we place a lid on the top, remove gravity, and exert another
force $f^y$.  Evidently unless $f^x/f^y = \epsilon/\delta$, a motion of the
soft mode would result in work, and the system would move.  Thus $f^y$ plus
the condition of no motion determines $f^x$.  This condition translates
into an imposed proportionality between the stresses $\sigma^{yy}$ and
$\sigma^{xx}$, as in the lattice example above.  The soft modes have
continuum consequences.
\section{Sequential packing under gravity}\label{sequential}
In the previous
section we have shown that a packing of frictionless spherical beads is an
anomalous solid from the point of view of classical elastic theory. The fact
that the average coordination number in such a packing is exactly $2d$ for the
infinite system  supports unidirectional, propagating stress. Now we elaborate
this concept in more detail, by deriving particular laws for microscopic and
macroscopic force transfer adopting a particular  packing procedure.   We
suppose that the beads are deposited one by one in the presence of gravity. The
weight of any  new sphere added to the  existing packing   must be balanced by
the reactions of the supporting beads. This is possible only if the number of
such supporting contacts is equal to the dimensionality $d$. Any larger number
of contacts requires a specific relationship between the sizes and coordinates
of the supporting beads, and  and thus occurs with vanishing probability. As a
result, the eventual packing has an average coordination number $2d$, like any
stable, frictionless pack.   In addition, it has a  further property: a partial
time-like ordering.  Namely, among any two contacting beads there is always one
which has found its place earlier than the other (the supporting one), and any
bead has exactly $d$ such  supporting neighbors. Note that the supporting bead
is not necessarily situated below the supported one in the geometrical sense.
The discussed ordering is  topological rather than spatial.
\par
One could expect that although any bead has exactly $d$ supporters at the
moment of deposition, this may
change later.  Specifically, adding an extra bead to the packing  may
result in the
violation of positivity of some contact force in the
bulk\cite{Claudin.PRL.1997}.
This will lead to a rearrangement of the network.   For the moment we
assume that the
topology of the sequential packing is preserved in the final state of the
system, and
return to the effect of rearrangements in Section \ref{reality}.
\par
  The  partial ordering of the  sequential packing considerably simplifies
the calculation of the  force distribution. Indeed, any force applied to a
bead can be  uniquely decomposed into the $d$  forces on the supporting
contacts. This means that the force balance enables us to determine all the
``outcoming" (downward) forces if the ``incoming" ones are known.
Therefore, there is a simple unidirectional procedure of determination of
all the forces in the system. Below, we use this observation to construct a
theory of stress propagation on the macroscopic scale.
\section{Mean-field stress}\label{macroscopic}
 We will characterize any inter-bead contact in a sequential packing with a
unit vector directed from the center of supported bead $\alpha$ toward the
supporting one $\beta$,
\begin{equation}
{\bf n}_{\alpha \beta}= \frac{{\bf x}_\beta - {\bf x}_\alpha }{\left|{\bf
x}_\beta - {\bf x}_\alpha\right|}
\end{equation}
 The stress distribution in the frictionless  packing is given if a
non-negative magnitude of the  force acting along any of the above  contact
unit vector is specified. We denote such scalar contact force as
$f_{\alpha\beta}$

\par
 The total force to be transmitted from some bead $\alpha$ to its
supporting neighbors is the sum of all the incoming and external (e.g.
gravitational) forces:

\begin{equation}{\bf F}_\alpha=({\bf f}_{\rm ext})_\alpha+\sum
_{\beta(\rightarrow\alpha)} {\bf n}_{\beta\alpha} f_
{\beta\alpha}\end{equation}

\par
  Here $\beta(\rightarrow
\alpha)$ denotes all the beads supported by $\alpha$. Since there are
exactly $d$ supporting contacts for any bead in a sequential packing, the
above force can be uniquely decomposed onto the corresponding $d$
components, directed along the  outcoming vectors ${\bf n}_
{\alpha\gamma}$. This gives the values of the outcoming forces.  The $f$'s
may be compactly expressed in terms of a generalized scalar product
$\expectation{...|...}_\alpha$:
\begin{equation}
f_ {\alpha\gamma}= \expectation{{\bf F_{\alpha}}|{\bf n}_
{\alpha\gamma}}_\alpha
\end{equation}
 The scalar product $\expectation{...|...}_\alpha$ is defined such that
$\expectation{{\bf n}_{\alpha \gamma}|{\bf n}_
{\alpha\gamma'}}_\alpha=\delta_{\gamma \gamma'}$. (all the Greek indices
count beads, not spatial dimensions). In general, it does not coincide with
the conventional scalar product.   If now some force is applied to certain
bead in the packing, the above projective procedure allows one to determine
the response of the system, i.e. the  change of the contact  forces between
all the beads below the given one. In other words one can follow how the
perturbation propagates downward. Since the equations of mechanical
equilibrium are linear, and beads are assumed to be rigid enough to
preserve their sizes, the response of the system to the applied force is
also  linear. This linearity can be destroyed only by violating the
condition of positivity of the contact forces, which implies the
rearrangement of the packing. While the topology (and geometry) of the
network is preserved, one can introduce the Green function to describe the
response of the system to the applied forces. Namely, force ${\bf
f}_\lambda$ applied to  certain bead $\lambda$ results in the following
additional force acting on another bead, $\mu$ (lying below $\lambda$):
\omitt{\begin{equation}
f_{\lambda \mu} = \expectation{{\bf f}_\lambda |
\widehat{G}_{\lambda\mu}}_\lambda
\end{equation}
}
\begin{equation}
\vector f_\mu = \widehat{\bf G}_{\mu \lambda} \cdot \vector f_\lambda
\end{equation}
\par
 Here $\widehat{\bf G}_{\lambda\mu}$ is a tensor Green function, which can
be calculated as the superposition of all the projection sequences (i.e.
trajectories), which lead from $\lambda$ to $\mu$.
\par

 The stress field $\sigma^{ij}$ in the system of frictionless spherical
beads can be introduced in the following way \cite{stress}:

\begin{equation}
\sigma^{ij}({\bf x})=\sum_{\alpha}\sum_{\beta(\leftarrow
\alpha)}f_{\alpha\beta} n_{\alpha\beta}^i n_{\alpha\beta}^j
R_{\alpha\beta}\delta({\bf x}_\alpha-{\bf x})\end{equation}

\par
  Here $R_{\alpha\beta}=\left|{\bf x}_\alpha - {\bf x}_\beta
\right|$. As  we have just shown, the magnitude of the force
$f_{\alpha\beta}$ transmitted along the contact unit vector ${\bf
n}_{\alpha\beta}$ can be expressed as an appropriate  projection of the
total force $F_\alpha$ acting on the bead $\alpha$ from above. This allows
one to express the stress tensor in terms of the vector field $F_\alpha$:
\begin{equation} \label{sigma.of.n}
\sigma^{ij}({\bf x})=\sum_{\alpha}\sum_{\beta(\leftarrow
\alpha)} \expectation{ F_{\alpha}| {\bf n}_{\alpha\beta}}_\alpha
n_{\alpha\beta}^i n_{\alpha\beta}^j R_{\alpha\beta}\delta({\bf
x}_\alpha-{\bf x})
\end{equation}
\par
 In order to obtain the continuous macroscopic description of the system,
one has to perform the averaging of the stress field over a region much
larger than a bead. At this stage we make a  mean-field approximation for
the force $\vector F_\alpha$ acting on a given bead from above: we replace
$\vector F_\alpha$ by its average $\overline{\vector F}$ over the region.
To be valid, this assumption requires that
\begin{eqnarray}
\sum_{\alpha \beta} \expectation{(\vector F_\alpha - \overline{\vector F}) |
{\bf n}_{\alpha \beta} }_\alpha
 {\bf n}^i_{\alpha \beta}  {\bf n}^j_{\alpha \beta} R_{\alpha \beta}
\nonumber\\
\muchlessthan
\sum_{\alpha \beta} \expectation{\overline{\vector F} |  {\bf n}_{\alpha
\beta} }_\alpha
 {\bf n}^i_{\alpha \beta}  {\bf n}^j_{\alpha \beta} R_{\alpha \beta}
\end{eqnarray}
\par
For certain simple geometries, the mean-field approximation is exact.  One
example is the simple square lattice treated in Section \ref{counting}.  In
any regular lattice with one bead per unit cell, all the $\vector
F_\alpha$'s must be equal under any uniform applied stress.  Thus replacing
$\vector F_\alpha$ by its average changes nothing.  If this lattice is
distorted by displacing its soft modes, the $\vector F_\alpha$ are no
longer equal  and the validity of the mean-field approximation can be
tested.  Figure \ref{rhombuses} shows a periodic distortion with fourbeads
per unit cell.  For example, under an applied vertical force, the bottom
forces oscillate to the left and right.  Nevertheless, the stress crossing
the bottom row, like that crossing the row above it, is the average force
times the length.  One may verify that the $\vector F_\alpha$ may also be
replaced by its average when the applied force is horizontal.  Though the
mean-field approximation is exact in these cases, it is clearly not exact
in all.  In the lattice of Figure \ref{rhombuses} the mean field
approximation may be inexact if one considers a region not equal to a whole
number of unit cells.

A disordered packing may be viewed as a superposition of periodic soft
modes like those of Figure \ref{rhombuses}.  Each such mode produces
fluctuating forces, like those of the example.  But after averaging over an
integer number of unit cells, the stress may depend on only the average
force $\overline{\vector F}$.  A disordered packing need not have a fixed
co-ordination number as our example does.  This is another possible source
of departure from the mean-field result.
\par
 Now, it becomes an easy task to perform a  local averaging of the  Eq.
\ref{sigma.of.n} for the stress field in the vicinity of a given point $\bf
x$, replacing $\vector F_\alpha$ by its average:
\begin{equation}
\overline{\sigma^{ij}}({\bf x})=\rho \overline{F^k}({\bf x})
\tau^{kij}({\bf x})
\label{constitutive}
\end{equation}
  Here $\rho$ is the bead density, $\overline{\bf F}({\bf x})$ is the force
${\bf F}_\alpha$ averaged over the beads $\alpha$ in the vicinity of the
point ${\bf x}$, and the third-order tensor $\hat\tau$ characterizes the
local geometry of the packing:
\begin{equation}\tau^{kij}({\bf x})= \overline{\ket{{\bf n}_{\alpha\beta}
 }^k_\alpha n_{\alpha\beta}^i n_{\alpha\beta}^j R_{\alpha\beta}}
\label{tau}
\end{equation}
This equation is similar in spirit to one derived by Edwards for the case
of a $d+1$ co-ordinated packing of spheres with
friction\cite{Edwards.les.Houches}.  Our geometric tensor $\tau$ plays a
role analogous to that of the fabric tensor in that treatment.
\par

\par
  The stress field satisfies the force balance equation, Eq.
({\ref{balance}).  Since this   is a vector equation, it normally fails to
give a complete description of the tensor stress field. In our case,
however, the stress field has been expressed in terms of the vector field
$\vector F$. This creates a necessary closure for the force balance
equation. It is important to note that the proposed macroscopic formalism
is complete for a system of arbitrary dimensionality: there is a  single
vector equation and a single vector variable. We now discuss the
application of the above macroscopic formalism in two special cases. First
we consider the equations of stress propagation in two dimensions. Then we
discuss a packing of arbitrary dimensionality but with uniaxial symmetry.
It is assumed to have no preferred direction other than that of gravity.

\par
\subsection {Two-dimensional packing.}
In two dimensions, according to Eq. (\ref{constitutive}), the stress tensor
$\hat \sigma$ can be written as a linear combination of two $\tau$ tensors.
\begin{equation}
\hat \sigma = F_{1}\hat \sigma_{1} + F_{2}\hat \sigma_{2},
\label{ATs20}
\end{equation} where $[\hat \sigma_{1}]^{ij} = \tau^{1ij}$ and $[\hat
\sigma_{2}]^{ij} = \tau^{2ij}$.  Since the $\hat \sigma_{1}$ and $\hat
\sigma_{2}$ are properties of the medium and are presumed known, the
problem of finding the stress profile $\hat \sigma(x)$ becomes that of
finding $F_{1}$ and $F_{2}$ under a given external load.  Rather than
determining these $F$'s directly, we may view Eq. (\ref{ATs20}) as a
constraint on $\hat
\sigma$.  The form (\ref{ATs20}) constrains $\hat\sigma$ to lie in a
subspace of the three-dimensional space of stress components $\vec \sigma
\equiv (\sigma^{xx}, \sigma^{yy}, \sigma^{xy})$.  It must lie in the
two-dimensional subspace spanned by $\vec \sigma_{1}$ and $\vec
\sigma_{2}$.  This constraint amounts to one linear constraint on the
components of $\sigma$, of the form

\begin{equation}
\sigma^{ij}u^{ij}=0
\label{null}
\end{equation}  where the $\hat u$ tensor is determined by
$\hat\sigma_{1}$, and  $\hat\sigma_{2}$.  Specifically, $\hat u$ may be
found by observing that the determinant of the vectors $\vec\sigma$,
$\vec\sigma_{1}$, $\vec\sigma_{2}$ must vanish.  Expanding the determinant
by minors to obtain the coefficients of the $\sigma^{ij}$, one finds
\begin{equation}\hat{u}=
\left(       \begin {array}{cc}               \left|
\begin {array}{cc}

\sigma_{1}^{yy} & \sigma_{2}^{yy}\\

\sigma_{1}^{xy} & \sigma_{2}^{xy}                     \end {array}

\right|               &              \left|

\begin {array}{cc}                     \sigma_{1}^{xx} &
\sigma_{2}^{xx}\\                     \sigma_{1}^{yy} &
\sigma_{2}^{yy}                     \end {array}              \right|

\\               &               \\               \left|

\begin {array}{cc}                     \sigma_{1}^{xx} & \sigma_{2}^{xx}\\
\sigma_{1}^{yy} & \sigma_{2}^{yy}                     \end {array}
\right| &  \left|                     \begin {array}{cc}
\sigma_{1}^{xy} & \sigma_{2}^{xy}\\                     \sigma_{1}^{xx} &
\sigma_{2}^{xx}                     \end {array}              \right|
\end{array}
\right)
\end{equation}

\par
Eq. \ref{null} has the same ``null-stress" form as that introduced by
Wittmer et al \cite{Wittmer.Nature}, whose original  arguments were based
on a  qualitative  analysis of the problem. By an appropriate choice of the
local co-ordinates ($\xi$, $\eta$),  the $\hat u$ tensor can be transformed
into  co-ordinates such that $u^{\xi\xi} = u^{\eta\eta} = 0$.  Then the
null stress condition becomes $\sigma^{\xi\eta} = \sigma^{\eta\xi} = 0$.
 This implies that, according to  force balance equation (\ref{balance}),
the non-zero diagonal components of the stress tensor ``propagate"
independently along the corresponding characteristics, $\xi=const$ and
$\eta=const$:
\begin{equation}
\partial^\xi \sigma^{\xi\xi}=f_{\rm ext}^\xi \quad \quad
\partial^\eta \sigma^{\eta\eta}=f_{\rm ext}^\eta
\end{equation}

Our microscopic approach gives an alternative foundation for the
null-stress condition, Eq. (\ref{null}), and allows one to relate the
tensor $\hat {u}$ in this equation to the local geometry of the packing.
Our general formalism is not limited to the two-dimensional case, and in
this sense, is a generalization of the null-stress approach.
\par
\subsection {Axially-symmetric packing.}
Generally, there are  two preferred directions in the sequential packing:
that of the  gravitational force, ${\bf g}$ , and that of the  growth
surface ${\bf n}$. In the case when these two directions coincide, the form
of the third-order tensor $\hat{\tau}$, Eq. (\ref{tau}), should be
consistent with the axial symmetry associated with the single preferred
direction, ${\bf n}$. Since $\tau^{kij}$ is symmetric with respect to
$i\leftrightarrow j$ permutation, it can be  only a linear  combination of
three tensors: $n^kn^in^j$, $n^k\delta^{ij}$ and
$\delta^{ki}n^j+\delta^{kj}n^i$,  for general spatial dimension $d$.

Let $\sigma^{ij}$ be the  stress tensor in the $d$-dimensional space
($i,j=0...d-1$, and index '0' corresponds to the vertical direction).  From
the point of view of rotation around the vertical axis the stress splits
into scalar $\sigma^{00}$, $d-1$ dimensional vector  $\sigma^{0 a}$
($a=1...d-1$)  $d-1$ dimensional tensor $\sigma^{a b}$. According to our
constitutive equation \ref{constitutive}, the stress should be linear in
vector $\bf F$, which itself splits into a scalar $F^0$ and a vector $F^a$
with respect to horizontal rotations. Since the  material tensor $\tau$ is
by hypothesis axially symmetric, the only way that the ``scalar''
$\sigma^{00}$ may depend on $\bf F$ is to be proportional to ``scalar''
$F^0$.  Likewise, the only way   ``tensor'' $\sigma^{a b}$ can be linear in
$\bf F$ is to be proportional to $\delta^{a b} F^0$. Therefore, in the
axially-symmetric case
\begin{equation}
\sigma^{a b}= \lambda  \delta^{a b}\sigma^{00},
\end{equation}
 where the constant $\lambda$ is \eg $\tau^{0 1 1}/\tau^{0 0 0}$.  This
constitutive equation allows one to convert the force balance equation
(\ref{balance}) to the following form: \omitt{TW bk 110 p 44}
\begin{equation}
\partial^0 \sigma^{00}+\partial^a \sigma^{a 0}=f^0_{ext};\indent
 \partial^0 \sigma^{a 0}+\lambda\partial^a \sigma^{00}=f^{a}_{ext}
 \end{equation}
 In the case of no external force,  we may take $\partial^0$ of the first
equation and combine with the second to yield a wave equation for
$\sigma^{0 0}$.  Evidently $\sigma^{a b}$, being a fixed multiple of
$\sigma^{0 0}$, obeys the same equation.  Similar manipulation yields the
same wave equation for $\sigma^{0 a}$ and $\sigma^{a 0}$.  Thus every
component of stress satisfies the wave equation  with vertical direction
playing the role of time and $\sqrt \lambda$ being the propagation
velocity.

\section{ Discussion}\label{reality}

In this section we consider how well our model should describe real systems
of rigid, packed units.  As stated above, our model is most relevant for
emulsions or dense colloidal suspensions, whose elementary units ae well
described as frictionless spheres.  Under very weak compression the forces
between such units match our model assumptions.  However, our artificial
procedure of sequential packing bears no obvious resemblance to the
arrangements in real suspensions.  We argue below that our model may well
have value even when the packing is not sequential.  More broadly we may
consider the connection between our frictionless model and real granular
materials with friction.  The qualitative effect of adding friction to our
sequential packing is to add constraints so that the network of contacts is
no longer  minimally connected.  Thus the basis for a null-stress
description of the force transmission is compromised.  We argue below that
friction should cause forces to propagate as in an elastic medium, not via
null-stress behavior.
\subsection{Sequential packing}
 We first consider the consequences of our sequential packing assumption.
One consequence is that each bead has exactly $d$ supporting contacts.
These lead successively to earlier particles, forming a treelike
connectivity from supported beads to supporters.  Although the counting
arguments of Section \ref{counting} show that the propagating stress
approach should be applicable to a wide class of frictionless systems, the
continuum description of Section \ref{macroscopic} depends strongly on the
assumed  sequential order.  Now, most packings are not sequential, and even
when beads are deposited in sequence, they may undergo rearrangements that
alter the network of connections.  However, it is possible to modify our
arguments to take account of such re-arrangements.  Our reasoning depends
on the existence of $d$ supporting contacts for each bead.  Further, every
sequence of supporting contacts starting at a given bead must reach the
boundary of the system without passing through the starting bead: there
must be no closed loops in the sequence.
\par
Even in a non-sequential packing we may define a network of supporting
contacts.  First we define a downward direction. Then, for any given bead
in the pack, we {\it define} the supporting contacts to be the $d$ lowest
contacts.  With probability 1, each bead has at least $d$ contacts.
Otherwise it is not stable.  Typically a supporting bead lies lower than
the given bead.  Thus the typical sequence of supporting contacts leads
predominantly downward, away from the given bead, and returns only rarely
to the original height.  A return to the original {\it bead} must be even
more rare.
One may estimate the probability that there is a loop path of supporting
contacts under simple assumptions about the packing.  As an example we
suppose the contacts on a given bead to be randomly distributed amongst the
12 sites of a randomly-oriented close-packed lattice.  We further imagine
that these sites are chosen independently for each bead, with at least one
below the horizontal.  Then the paths are biased random walks with a mean
steplength of .51 \omitt{bk 110 p. 46} diameters and a root-mean-square
steplength of about 1.2  times the mean.  The probability of a net upward
displacement of 1 or more diameter is about one percent.  It appears that
our neglect of loop paths is not unreasonable.
\subsection{Friction}
 The introduction of friction strongly affects most of our arguments.
Friction creates transverse as well as normal forces at the contacts.
  The problem is to determine positions and orientations of the beads that
lead to balanced forces and torques on each.  If the contact network is
minimally connected, the forces can be determined without reference to
deformations of the particle.  But if the network has additional
constraints, it is impossible to satisfy these without considering
deformation.  This is no less true if the beads are presumed very rigid.
We first give an example to show that in a generic packing the
deformability alters the force distribution substantially.  We then give a
prescription for defining the deformation and hence the contact forces
unambiguously.
\par
In our example we imagine a two-dimensional sequential packing and focus on
a particular bead, labeled 0, as pictured in Figure \ref{threebeads}.  We
presume that the beads are deposited gently, so that each contact forms
without tangential force.  Thus when the bead is deposited, it is minimally
connected: its weight determines the two  (normal) supporting forces,
labeled 1 and 2.  Thenceforth no slipping is allowed at the contact.  Later
during the construction of the pack bead 0 must support the force from some
subsequent bead.  This new force is normal, since it too arises from an
initial contact.  But the new force creates tangential forces on the
supporting contacts 1 and 2.  To gauge their magnitude, we first suppose
that there is no friction at contacts 1 and 2, while the supporting beads
remain immobile.  Then the added force $F$ leads to a compression.  We
denote the compression of the contact 1 as $\delta$.  With no friction, the
contact 2 would undergo a slipping displacement by an amount of order
$\delta$.  Friction forbids this slipping and decrees deformation of the
contact instead.  The original displacement there would create an elastic
restoring force of the same order as the original $F$.  Thus the imposition
of friction creates new forces whose strength is comparable to those
without friction.  The frictional forces are not negligible, even if the
beads are rigid. Increasing the rigidity lessens the displacements $\delta$
associated with the given force $F$, but it does not alter the ratio of
frictional to normal forces.
Neither are the frictional forces large compared to the normal forces.
Thus a coefficient of friction $\mu$ of order unity should be sufficient to
generate enough frictional force to prevent slipping of a substantial
fraction of the contacts.
\par
The contact forces $T_1$ and $T_2$ cannot be determined by force balance
alone, as they could in the frictionless case.  Now the actual contact
forces are those which minimize the elastic energy of deformation near the
two contacts.  This argument holds not just for spheres but for general
rounded objects.
\par
Though the new tangential forces complicate the determination of the
forces, the determination need not be ambiguous.  We illustrate this point
for a sequential packing on a bumpy surface with perfect friction.  We
choose a placement of the successive beads so that no contact
re-arrangements occur.  If only a few beads have been deposited in the
container, the forces are clearly well determined.  Further, if the forces
are presumed well-determined up to the $M$th bead, they remain so upon
addition of the $M+1$st bead.  We presume as before that the new bead
exerts only normal forces on its immediate supporters.  Each supporter thus
experiences a well-defined force, as shown in Section \ref{counting}.  But
by hypothesis, these supporting beads are part of a well-connected, solid object, whose contacts may be regarded as fastened together.  Thus the displacem
ents and rotations of each bead are a well-defined function of any small
applied load.  Once the $M+1$'st bead has been added, its supporting
contacts also support tangential force, so that it responds to future loads
as part of the elastic body.
\par We conclude that a sequential packing with perfect friction, under
conditions that prevent contact rearrangements, transmits forces like a
solid body.  Small changes in stress $\delta \sigma^{ij}$ in a region give
rise to proportional changes in the strain $\delta \gamma^{k \ell}$.  This
proportionality is summarized by an elasticity tensor $K^{ijk\ell}$:
$\delta \sigma^{ij} = K^{ijk\ell} \delta \gamma^{k\ell}$.  The elastic
tensor $K$ should depend in general on how the pack was formed; thus it may
well be anisotropic.
\par
 This elastic picture is compromised when the limitations of friction are
taken into account.   As new beads are added, underlying contacts such as
contacts 1 and 2 of Figure \ref{threebeads} may slip if the tangential
force becomes too great.   Each slipping contact relaxes so as to satisfy a
fixed relationship between its normal force $N$ and its tangential force
$T$: \viz $|T| = \mu |N|$.  If $\mu$ were very small, virtually all the
contacts would slip until their tangential force were nearly zero.  Then
the amount of stress associated with the redundant constraints must become
small and the corresponding elastic moduli must become weak.  Moreover, as
$\mu$ approaches 0, the material on any given scale must become difficult
to distinguish from a frictionless material with unidirectional stress
propagation.  Still, redundant constraints remain on the average and thus
the ultimate behavior at large length scales (for a given $\mu$) must be
elastic, provided the material  remains
homogeneous.
\par

\subsection{Force-generated contacts}
Throughout the discussion of frictionless packs we have ignored geometric
configurations with probability zero, such as beads with redundant
contacts. Such contacts occur in a close-packed lattice of identical disks,
for example.  Though such configurations are arbitrarily rare in principle,
they may nevertheless play a role in real bead packs.  Real bead packs have
a finite compressibility; compression of beads can create redundant
contacts.  Thus for example a close-packed lattice of identical spheres has
six contacts per bead, but if there is a slight variability in size, the
number of contacts drops to four.  The remaining two beads adjacent to a
given bead do not quite touch.  These remaining beads can be made to touch
again if sufficient compressive stress is applied.  Such stress-induced
redundant contacts must occur in a real bead with some nonzero density
under any nonzero load.  These extra contacts serve to stabilize the pack,
removing the indeterminate forces discussed in Section \ref{counting}.
To estimate the importance of this effect, we consider a large bead pack
compressed by a small factor $\gamma$.  This overall strain compresses a
typical contact by a factor of order $\gamma$ as well.  The number of new
contacts may be inferred from the pair correlation function $g(r)$.  Data
on this $g(r)$ is available for some computer-generated sequential packings
of identical spheres of radius $R$\cite{Pavlovitch}.  These data show that
$g(r)$ has a finite value near 1 at $r=2R$.  Thus the number of additional
contacts per bead that form under a slight compression by an amount $\delta
r$ is given by $\delta \overline{Z} = 6\phi g(2R)\delta r/R \aboutequal 4
\gamma$.  Here $\phi\aboutequal .6$ is the volume fraction of the beads.
These extra contacts impose constraints that reduce the number of
undetermined boundary forces in a compact region containing $M'$ beads and
roughly $M'^{2/3}$ surface beads.  The remaining number of undetermined
boundary forces now averages $\half N_{\rm ext} - M'\delta \overline{Z}$.
The first term is of order $M'^{2/3}$, and must thus be smaller than the
second term for $M'^{1/3} \aboutequal (\delta\overline{Z})^{-1}$.  For $M'$
larger than this amount, there  are no further undetermined forces and the
region becomes mechanically stable.
Moukarzel\cite{Moukarzel.PRL} reaches a similar conclusion by a somewhat
different route.
\par
If the pack is compressed by a factor of $\gamma$, stability occurs for
$M'^{1/3} \greaterthanorabout 1/\gamma$---a region roughly $1/\gamma$ beads
across.  In a typical experiment \cite{Nagel.Liu} the contact compression
$\gamma$ is $10^{-4}$ or less, and the system size is far smaller than
$10^4$ beads.
Thus compression-induced stability should be a minor effect here.
Still, this compression-induced stability might well play a significant
role for large and highly compressed bead packs such as compressed
emulsions\cite{Langer.Liu}.  In some of the large packs of Ref.
\cite{load}, compression-induced stability may also be important.

\subsection{Experimental evidence}

We have argued above that undeformed, frictionless beads should show
unidirectional, propagating forces while beads with friction should show
elastic spreading of forces.  The most direct test of these contrasting
behaviors is to measure the response to a local perturbing force
\cite{Claudin.PRL.1997}.
 Thus \eg if the pile of Figure \ref{elastic.vs} is a null-stress medium,
the local perturbing force should propagate along a light cone and should
thus be concentrated in a ring-like region at the
bottom\cite{Bouchaud.Cates.PRL}.  By contrast, if the pile is an elastic
medium the perturbing force should be spread in a broad pattern at the
bottom, with a maximum directly under the applied force.
Existing experimental information seems inadequate to test either
prediction, but experiments to measure such responses are in progress
\cite{response.experiments}.

As noted above, emulsions and colloids are good realizations of the
frictionless case.  The contacts in such systems are typically organized by
hydrostatic pressure or by flow, rather than by gravity.  Still, our
general arguments for unidirectional propagation should apply.  Extensive
mechanical measurements of these systems have been made
\cite{Weitz.Mason,VanderWerff}.  The shear modulus study of Weitz and Mason
\cite{Weitz.Mason} illustrates the issues.  The study spans the range from
liquid-like behavior at low volume fractions to solid-like behavior at high
volume fractions.  In between these two regimes should lie a state where
the emulsion droplets are well connected but little deformed.  The emulsion
in this state should show unidirectional force transmission.  It is not
clear how this should affect the measured apparent moduli.

Other indirect information about force propagation comes from the load
distribution of a granular pack on its container, such as the celebrated
central dip under a conical heap of sand \cite{load}.  These data amply
show that the mechanical properties of a pack depend on how it was
constructed.  Theories postulating null-stress behavior have successfully
explained these data \cite{Wittmer.Nature}.  But conventional
elasto-plastic theories have also proved capable of producing a central dip
\cite{Goddard}.  An anisotropic elastic tensor may also be capable of
explaining the central dip.

Another source of information is the statistical distribution of individual
contact forces within the pack or at its boundaries.  The measured forces
become exponentially rare for strong
forces\cite{Coppersmith.etal,Mueth}.  Such exponential falloff is predicted by
Coppersmith's ``q model" \cite{Coppersmith.etal}, which postulates
unidirectional force propagation.  Still, it is not clear whether this
exponential falloff is a distinguishing feature of unidirectional propagation.
A disordered elastic material might well show a similar exponential
distribution.

Computer simulations should also be able to test our predictions.  Recent
simulations\cite{Thornton,Langer.Liu} have focussed on stress-induced
restructuring of the force-bearing contact network.  We are not aware of a
simulation study of the transmission of a local perturbing force.  Such a
perturbation study seems quite feasible and would be a valuable test.  We have
performed a simple simulation to test the mean-field description of stress in
frictionless packs.  Preliminary results agree well with the predictions.  An
account of our simulations will be published separately.

\section{Conclusion}
In this study we have aimed to understand how force is transmitted in
granular media, whether via elastic response or via unidirectional
propagation.  We have identified a class of disordered systems that ought
to show unidirectional propagation.  Namely, we have shown that in a general
case a  system of frictionless rigid particles must be isostatic, or
minimally connected. That is, all the
inter-particle forces can in principle be determined from the force balance
equations.  This contrasts with statically undetermined, elastic
systems, in which the forces cannot be determined without
self-consistently finding the displacements induced by those forces.  Our
general equation-counting arguments suggest that isostatic property  of the
frictionless packing results in the unidirectional propagation of the
macroscopic stress.

We were able to demonstrate this unidirectional propagation explicitly by
specializing to the case of sequential packing.  Here the stress obeys a
condition of the previously postulated null-stress form\cite{Wittmer.Nature};
our system provides a microscopic justification for the null-stress postulate.
Further, we could determine the numerical coefficients entering the null-stress
law from statistical averages of the contact angles by using a mean field
hypothesis (decoupling Anzatz).  We have devised a numerical simulation to test
the adequacy of the sequential packing assumption and the mean-field
hypothesis.  The results will be reported elsewhere.

If we add friction in order to describe macroscopic granular packs more
accurately, the packing of rigid particles no longer needs to be isostatic, and
the  system is expected to revert to elastic behavior.  This elasticity
does not
arise from softness of the beads or from a peculiar choice of contact network.
It arises because contacts that provide only minimal constraints when created
can provide redundant constraints upon further loading.

We expect our formalism to be useful in understanding experimental granular
systems.  It is most applicable to dense colloidal suspensions, where static
friction is normally negligible.  Here we expect null-stress behavior to emerge
at scales large enough that the suspension may be considered uniform.  We
further expect that our mean-field methods will be useful in estimating the
coefficients in the null-stress laws.  In macroscopic granular packs our
formalism is less applicable because these packs have friction.  Still, this
friction may be small enough in many situations that our picture remains
useful.  Then our microscopic justification  may account for the practical
success of the null-stress postulate\cite{Wittmer.Nature} for these systems.

\section*{Acknowledgement}
The authors are grateful to the Institute for Theoretical Physics in Santa
Barbara,  for hospitality and support that enabled this research to be
initiated.  The authors thank the participants in the Institute's program on
Jamming and Rheology for many valuable discussions.  Many of the ideas reported
here had their roots in the floppy network theory of Shlomo Alexander.  The
authors dedicate this paper to his memory.  This work was supported in part by
the National Science Foundation under Award numbers PHY-94 07194, DMR-9528957
and DMR 94--00379.

\newpage \section*{Figures}
\begin{figure}
 \epsfxsize=\hsize $$\epsfbox{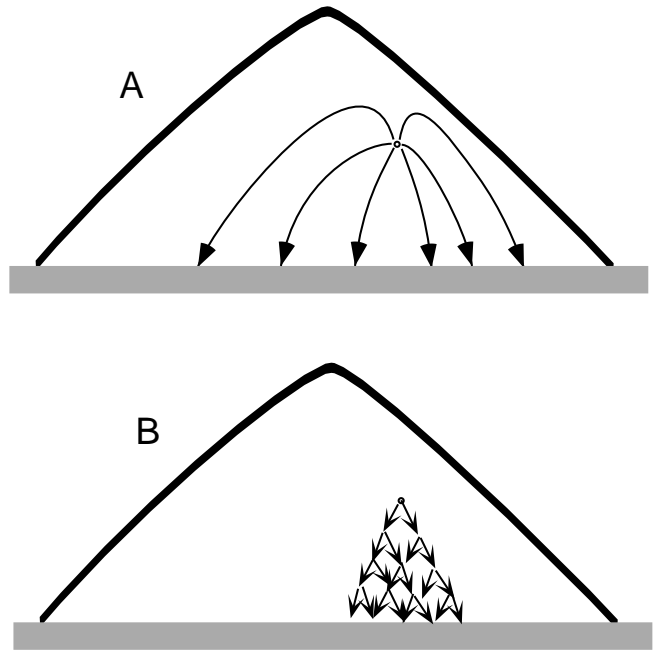}$$
\caption{
Contrast between elastic (elliptical) and unidirectional (hyperbolic) force
propagation in a sand-pile shaped object.
A: elastic response to an imposed local downward force.  Pictured lines of
force
represent the current of vertical force, \ie the stress contracted into a
downward unit vector.
Near the source, this field is symmetric about a horizontal plane through
the object
\protect\cite{Oseen}; part of the force is transmitted through points above the
source.  B: unidirectional response to an imposed local downward force.  The
imposed force is transmitted to neighbors below the source, and is further
transmitted to neighbors below these.  No force is transmitted to points above
the source.}
\label{elastic.vs}
\end{figure}
\begin{figure}
 \epsfxsize=\hsize $$\epsfbox{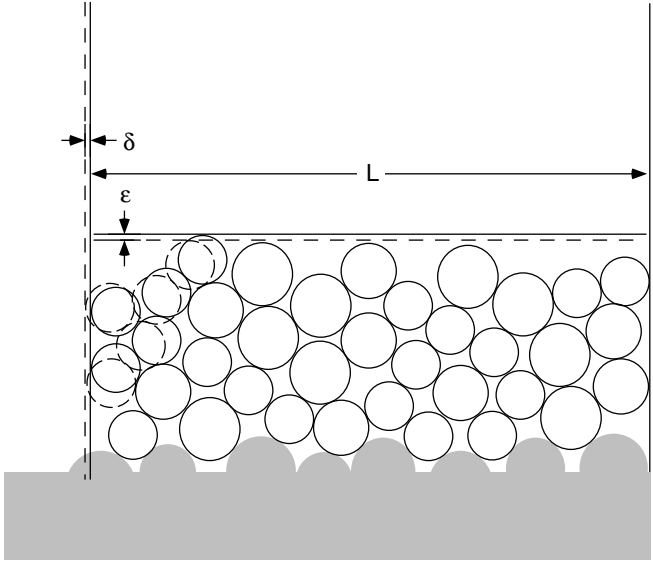}$$
\caption{
A sequential packing in a channel.  When left wall is displaced outward by
a small amount $\delta$, the beads shift to the positions shown by dashed
lines, and the top of the pile shifts by an amount $\epsilon$.}
\label{channel}
\end{figure}
\begin{figure}
 \epsfxsize=\hsize $$\epsfbox{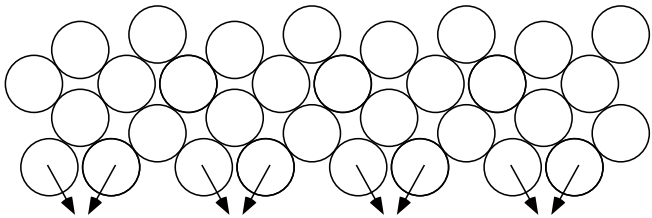}$$
\caption{
A buckled square lattice illustrating the propagation of inhomogeneous
forces.  Bottom row of sites has alternating wide and narrow spacing.
Arrows indicate the unequal forces on these sites.}
\label{rhombuses}
\end{figure}
\begin{figure}
 \epsfxsize=\hsize $$\epsfbox{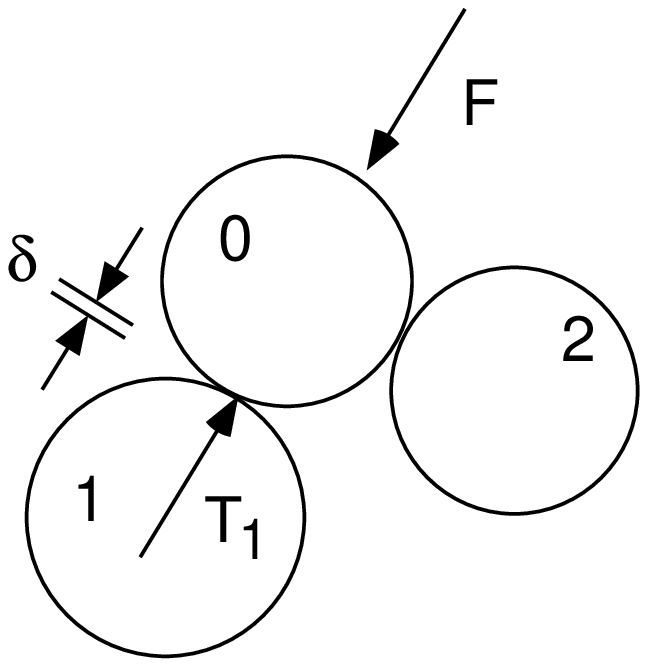}$$
\caption{
The effect of friction on a triad of beads.  In the absence of friction,
the applied force $F$ is transmitted entirely to contact 1, causing a
displacement $\delta$.
This would result in a sliding displacement of contact 2 by an amount
$\delta$.  With friction, contact 2 cannot slide; it must deform the
contact region by an amount of order $\delta$.  Thus the applied force $F$
is
shared between contacts 1 and 2.  The force is distributed so as to
minimize the total
elastic energy at contacts 1 and 2.}
\label{threebeads}
\end{figure}
}

\begin{thebibliography}{9}
\bibitem{Guyon.1990} For a review, see E. Guyon, S. Roux, A. Hansen, D.
Bideau, J.-P. Troadec and H. Crapo, {\sl Rep. Prog. Phys.} \vol{53} 373
(1990).
\bibitem{Grassberger} {\it Friction, Arching, Contact Dynamics} edited by
D. E. Wolf and P. Grassberger (Singapore, World Scientific, 1997).
\bibitem{load} T. Jotaki and R. Moriyama, {\em J. Soc. Powder Technol. Jpn}
{\bf 60}, 184 (1979); J. Smid and J. Novosad in Proc. of 1981 Powtech
Conf., {\sl Ind.Chem. Eng. Symp.} \vol{63}, D3V 1-12 (1981); \omitt{Contact
force distribution beneath a three-dimensional granular pile [an
experiment]}R. Brockbank, J. M. Huntley and R. C. Ball, J. Physique II, 7,
1521 (1997).
\bibitem{Edwards.Mounfield}\omitt{A theoretical model for the stress
distribution in granular matter .1. Basic equations}S. F. Edwards, C. C.
Mounfield {\sl Physica A} \vol{226} (1996).
\bibitem{Goddard} \omitt{Computations of dilatancy and yield surfaces for
assemblies of rigid frictional spheres} J. D. Goddard and A. K. Didwania
{\sl Quarterly J. Mech. Appl. Math.} \vol{51} 15, (1998); F. Cantelaube and
J. D. Goddard {\it Powders and Grains 97} (R. P. Behringer and J. T.
Jenkins, eds (Rotterdam, Balkema  1997) p. 185
\bibitem{Wittmer.Nature} J. P. Wittmer, P. Claudin, M. E. Cates and J. P.
Bouchaud {\sl Nature} \vol{382} 336 (1996); J. P. Wittmer, M. E. Cates and
P. Claudin, \omitt{ Stress propagation and arching in static sandpiles}
{\sl J. Phys. I France} {\bf 7}, 39 (1997).
\bibitem{Lacasse.Grest.Levine} Lacasse MD, Grest GS, Levine D
\artitle{Deformation of small compressed droplets} {\sl Phys Rev E}
\vol{54} 5436 (1996)
\bibitem{Langer.Liu} ~\omitt{Effect of Random Packing on Stress Relaxation
in Foam} S. A. Langer, Andrea J. Liu, {\sl J. Phys. Chem. B} \vol{101},
8667-8671 (1997)\par
\bibitem{Farr} R. S. Farr, J. R. Melrose, and R. C. Ball \omitt{Kinetic
theory of jamming in hard-sphere startup flows} {\sl Phys. Rev. E.}
\vol{55} 7203 (1997).
\bibitem{Luding.frictionless} S. Luding, \artitle{Stress Distribution in
static two dimensional granular model media in the absence of friction}
{\sl Phys. Rev. E} \vol{55} 4720 (1997)
\bibitem{Ball.Edwards.counting} S. F. Edwards, \omitt{The equations of
stress in a granular material} Physica A 249, 226 (1998).
\bibitem{Moukarzel.PRL} C. F. Moukarzel \omitt{Isostatic phase transition
and instability in stiff granular materials} cond-mat/9803120 {\sl Phys.
Rev. Lett.} \vol{81} 1634  (1998); \omitt{Granular matter instability: a
structural rigidity point of view} proceedings of {\it Rigidity theory and
applications} Traverse City MI, June 14-18 1998, Fundamental Material
Science Series, Plenum \omitt{<cristian@if.uff.br>}
\bibitem{Oron.Herrmann}G. Oron and H. J. Herrmann \artitle{Exact
calculation of force networks in granular media} {\sl Phys. Rev. E}
\vol{58} 2079 (1998).
\bibitem{Claudin.PRL.1997} \omitt{cited as discussing propagation of local
response} P. Claudin and J. P. Bouchaud \omitt{Static avalanches and giant
stress fluctuations in silos} {\sl Phys. Rev. Lett.} \vol{78} 231 (1997); P.
Claudin, J.-P. Bouchaud, M. E. Cates and J. P. Wittmer \omitt{models of stress
fluctuations in granular media, cond-mat/9710100,} {\sl Phys. Rev. E} \vol{57}
4441 (1998). \omitt{backup paper for \cite{Claudin.PRL.1997}}
\bibitem{Bouchaud.Cates.PRL} M. E. Cates, J. P. Wittmer, J. P. Bouchaud,
and P. Claudin \omitt{Jamming, force chains and fragile matter}{\sl Phys.
Rev. Lett.} \vol{81} 1841 (1998).
\bibitem{Oseen} \omitt{Text book deriving the Oseen tensor for
incompressible elastic medium.} J. N. Goodier, {\sl Phil. Mag.} \vol{22}
(1936) 678; H. M. Smallwood, {\sl J. Appl. Phys.} \vol{15} (1944) 758; E.
Guth, {\sl J. Appl. Phys.} \vol{16} (1945) 20; Z. Hashin,  {\it Proc. 4th
Internat. Congr. Rheol} vol. 3, E. H. Lee, ed., (Interscience, New York,
1965) p. 30. \omitt{The Stokes equation describing incompressible flow
around any object is the same as the equation for equilibrium of an
incompressible elastic medium around the same object.}
\bibitem{Coppersmith.etal} \omitt{Model for force fluctuations in bead
packs} S. N. Coppersmith, C-H Liu, S. Majumdar,O. Narayan, T. A. Witten
{\sl Phys. Rev. E.} \vol{53} 4673 (1996)
\bibitem{Socolar} \omitt{Average stresses and force fluctuations in noncohesive
granular materials} J. E. S. Socolar {\sl Phys. Rev. E} \vol{57} 3204 (1998).
\bibitem{Clement97} E. Cl\'ement and C. Eloy and J. Rajchenbach and J.
Duran, Stochastic
\omitt{Aspects of the Force Network in a Regular Granular Piling,} in {\it
Lectures on stochastic dynamics}, eds. T. P\"oschel and L. Schimanski-Geier,
(Heidelberg Springer 1997).
\bibitem{Mueth} \omitt{Force distribution in a granular medium} D. M. Mueth, H.
M. Jaeger and S. R. Nagel {\sl Phys. Rev. E.} \vol{57} 3164 (1998)
\bibitem{Rajdai96} F. Radjai and M. Jean and J. J. Moreau and S. Roux,
\omitt{Force Distribution in Dense Two-Dimensional Granular Systems,} {\sl
Phys.
Rev. Lett.} \vol{77} 274 (1996).
\bibitem{floppy.networks} S. Alexander, \omitt{AMORPHOUS
SOLIDS: THEIR STRUCTURE, LATTICE DYNAMICS AND ELASTICITY} {\sl Physics Reports}
\vol{296} 65 (1998).
\bibitem{Edwards.les.Houches} \omitt{Sam Edwards's mean-field formulation
of frictional packing with a fixed co-ordination number using fabric tensor
is this also Ball? Search of edwards articles yields nothing involving
fabric tensor.  I think this is Ball's work.}  R. C. Ball, private
communication September, 1997; R. C. Ball and S. F. Edwards, in
preparation; D. V. Grinev and R. C. Ball, in preparation.
\bibitem{VanderWerff}\omitt{LINEAR VISCOELASTIC BEHAVIOR OF DENSE
HARD-SPHERE DISPERSIONS} J. C. VanderWerff, C. G. Denkruif, C. Blom, and J.
Mellema, {\sl Phys. Rev. A} \vol{39} 795  (1989).
\bibitem{Watanabe.shear.thick}  Watanabe H, Yao ML, Osaki K, Shikata T,
Niwa H, Morishima Y \artitle{Nonlinear rheology of a concentrated spherical
silica suspension .2. Role of strain in shear-thickening} {\sl Rheologica
Acta} \vol{36} (1997); Nonlinear rheology and flow-induced structure in a
concentrated spherical silica suspension, {\sl Rheol Acta } \vol{37} 1-6
(1998)
\bibitem{Weitz.Mason} T. G. Mason, J. Bibette, and D. A. Weitz
\omitt{ELASTICITY OF COMPRESSED EMULSIONS} {\sl Phys. Rev. Lett.} \vol{75}
2051 (1995).
\bibitem{stress} L. D. Landau and E. M. Lifshitz, {\it Theory of
Elasticity} 3rd edition (Oxford, Pergamon 1986) Section 2.
\bibitem{Pavlovitch}\omitt{RANDOM PACKINGS OF SPHERES BUILT WITH SEQUENTIAL
MODELS} R. Jullien, A. Pavlovitch, and P. Meakin, {\sl J. Phys. A} \vol{25}
4103 (1992)
\bibitem{Nagel.Liu} C-H. Liu and S. R. Nagel \omitt{SOUND IN A GRANULAR
MATERIAL - DISORDER AND NONLINEARITY} {\sl Phys. Rev. B.} \vol{48} 15646
(1993)
\bibitem{response.experiments} S. Nagel, private communication; J.-P.
Bouchaud, private communication.
\bibitem{Thornton}\omitt{\?{this is not listed on his group web page or in
any index I can find} C. Thornton, in KONA Powder and Particle 15 (1997);}
C. Thornton and J. Lanier \omitt{Uniaxial compression of granular media :
numerical simulations and physical experiment}{\it Powders and Grains 97}
(R. P. Behringer and J. T.  Jenkins, eds (Rotterdam, Balkema  1997) p. 223
\bibitem{Behringer}\omitt{Stress fluctuations for continuously sheared
granular materials} B. Miller, C. O'Hern and R. P. Behringer {\sl Phys. Rev.
Lett.} \vol{77} 3110 (1996)  \end{thebibliography}
\end{document}